%% file: main.tex
\documentclass[%
 reprint,
superscriptaddress,
floatfix,
 amsmath,amssymb,
 aps,
 prx,
longbibliography
]{revtex4-2}

\usepackage{graphicx}
\usepackage{dcolumn}
\usepackage{bm}
\usepackage{placeins}
\usepackage{float}
\usepackage{hyperref}
\usepackage{placeins}
\usepackage{times}
\usepackage{physics}
\usepackage{gensymb}
\usepackage{xcolor}
\usepackage{float}
\usepackage{hyperref}
\usepackage{enumitem}
\usepackage[running]{lineno}
\usepackage{amsmath}
\usepackage{chemformula}
\usepackage{siunitx}
\hypersetup{
    colorlinks,
    citecolor=blue,
    filecolor=black,
    linkcolor=black,
    urlcolor=black
}
\usepackage[title,toc]{appendix}
\usepackage[T1]{fontenc}

\renewcommand{\thesubsection}{\arabic{subsection}.}

\newcommand\eryso{\ch{^{167}Er^{3+}:Y2SiO5} }

\begin{document}
\title{Integrated Dual-Resonator Architecture for Telecom Photon-Memory Entanglement}

\author
{\normalsize{Alexander Kolar$^{1,\dag}$, 
Ian Chin$^{1,\dag}$,
Conner Fong$^{1}$, 
Daniil M. Lukin$^{2}$,
Melissa A. Guidry$^{2}$,
Milan Palei$^{1}$,
Jelena Vu\v{c}kovi\'{c}$^{2}$,
Tian Zhong$^{1\ast}$}\\
\vspace{0.3cm}
\small{$^{1}$Pritzker School of Molecular Engineering, University of Chicago, Chicago, IL 60637, USA}\\
\small{$^{2}$E. L. Ginzton Laboratory, Stanford University, Stanford, CA 94305}\\
\small{$^\dag$These authors contributed equally to this work} \\
\small{$^\ast$tzh@uchicago.edu}
}



\begin{abstract}

Scalable quantum networks require the efficient generation and storage of entanglement between photonic qubits and quantum memories. Quantum repeaters based on absorptive rare-earth-ion photonic memories offer a promising route toward highly multiplexed quantum networking, but unifying spectrally matched photon sources and quantum memories within a common architecture remains a major challenge. Here we demonstrate an integrated photonic architecture for telecom photon-memory entanglement generation based on dual self-similar silicon carbide microring resonators. Connected by a fiber link, one resonator operates as an entangled photon-pair source, while the other functions as a cavity-enhanced atomic-frequency-comb quantum memory. The memory resonator reaches an ensemble cooperativity of 1.9 after hyperfine initialization and is spectrally matched to the source, enabling storage of entangled telecom photons without spectral modification. We generate and characterize photon-memory entanglement from a quantum interference visibility preserved before and after storage. Harnessing the strong source correlations and the high multimode capacity of the memory, we access high-dimensional entanglement spanning 63 temporal modes, reaching a maximum photon information efficiency of 5.1 Ebits per detected photon and a peak on-chip photon-memory entanglement rate of 5.6 kEbits per second. These results establish a route toward chip-scale quantum networking hardware operating over telecommunications infrastructure.

\end{abstract}

\maketitle
Quantum repeaters provide a promising avenue towards large-scale quantum networks by enabling the distribution of entanglement over distances beyond the limits imposed by optical loss~\cite{Kimble2008,Wehner2018}. A key requirement for repeater architectures is the ability to efficiently generate, store, and synchronize entanglement across many temporal and spectral modes~\cite{simon2007quantum,Sinclair2014,Zhong2015}. Achieving this capability requires quantum memories that can be seamlessly integrated with high-brightness entangled-photon sources and preferably operate at telecommunications wavelengths. Despite significant advances in both integrated photon-pair generation~\cite{clemmen2009continuous, zhang2017monolithic, lukin20204h, jiang2023quantum} and integrated photonic quantum memories~\cite{saglamyurek2011broadband,zhong2017nanophotonic,craiciu2019nanophotonic,meng2026efficient}, these technologies have largely developed independently, and no single platform has yet unified both functionalities within a common photonic architecture.

A central challenge in combining photonic sources and memories is spectral matching. Quantum memories generally operate over narrow optical bandwidths, whereas photon-pair sources often produce photons with substantially broader spectra, necessitating spectral filtering, cavity engineering, or frequency conversion that reduce entanglement generation rates and increase system complexity~\cite{jiang2023quantum,lago2021telecom,liu2021heralded,rakonjac2023transmission}. At the same time, both photon generation and quantum storage benefit from optical resonators: cavities enhance nonlinear optical interactions for efficient photon-pair generation~\cite{clemmen2009continuous,zhang2017monolithic,lukin20204h} while simultaneously enhancing light-matter coupling and memory performance~\cite{zhong2017nanophotonic,craiciu2019nanophotonic,meng2026efficient}. These benefits point to a natural architecture in which a common resonator platform can simultaneously address the requirements of both source and memory subsystems.

Here we demonstrate an integrated photonic architecture that combines telecom-band entangled-photon generation and quantum storage within a unified on-chip platform. The system consists of two self-similar silicon carbide (SiC) microring resonators that function as a cavity-enhanced spontaneous four-wave mixing (SFWM) photon-pair source and a cavity-enhanced atomic frequency comb quantum memory based on \eryso, respectively. Because the two devices share an identical design and fabrication process, the source and memory are spectrally matched, eliminating the need for spectral filtering, bandwidth modification, or quantum frequency conversion that commonly limit source-memory integration. The resulting resonator-enhanced interface simultaneously supports efficient photon-pair generation and strong collective light-matter interaction, yielding an ensemble coupling cooperativity of 1.9 and enabling efficient storage of telecom photons over a 200 MHz bandwidth. Using this architecture, we demonstrate high fidelity quantum storage of telecom entangled photons and high-dimensional photon-memory entanglement distributed across 63 temporal modes. The combination of designed source-memory matching and multimode storage enables an on-chip photon-memory entanglement generation rate up to 5.6 kEbits s$^{-1}$ and a photon information efficiency up to 5.1 Ebits per photon. These results establish a scalable platform for memory-enabled quantum networking over existing telecommunications fiber infrastructure and provide a practical path for integrated quantum interconnects.

\section*{Results}

\begin{figure*}
    \centering
    \includegraphics[width=\textwidth]{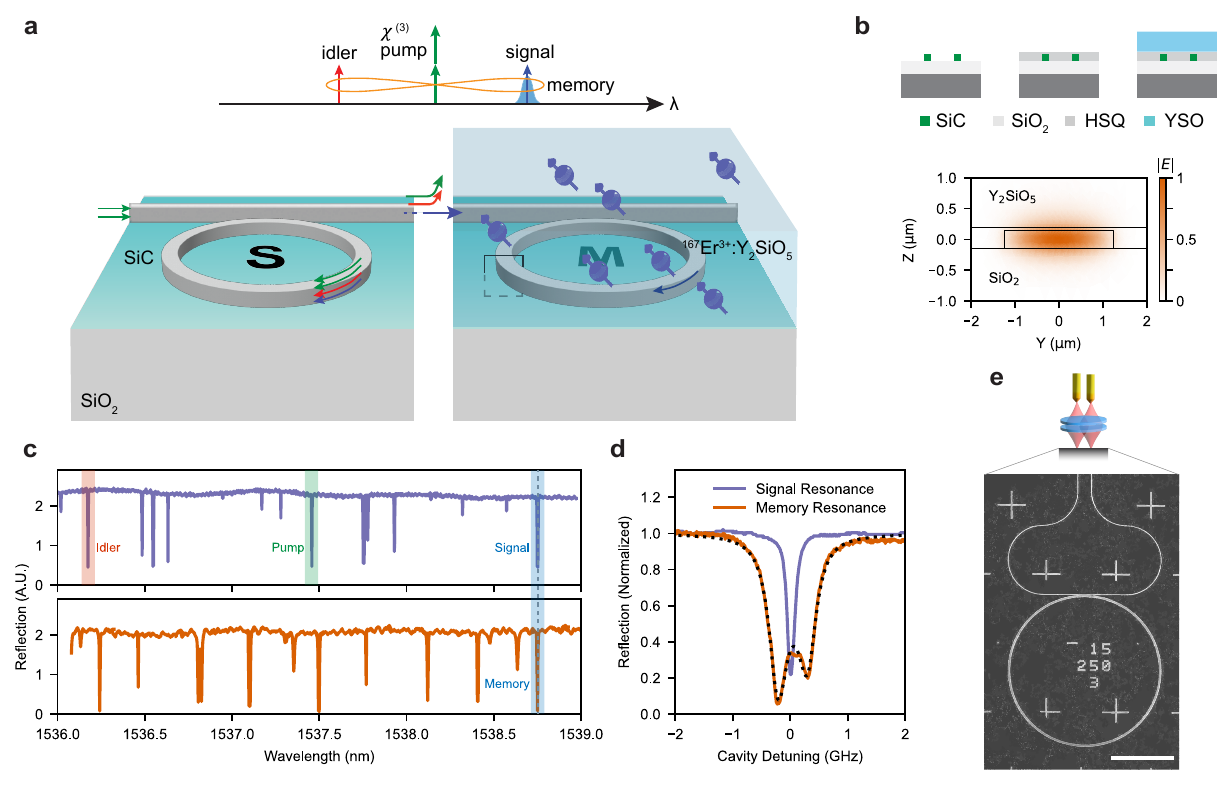}
    \caption{\textbf{Dual ring resonator architecture.} \textbf{(a)} Two identically fabricated ring resonators in silicon carbide are used to produce and store entangled photons. The first resonator produces non-degenerate signal (blue) and idler (red) photons through a cavity-enhanced spontaneous four-wave mixing process. The second resonator is evanescently coupled to an ensemble of \ch{^{167}Er^{3+}} ions in \ch{Y2SiO5}.
    \textbf{(b)} Bonding process to couple the erbium ensemble with the silicon carbide photonic circuits. After fabrication, a thin layer of HSQ is spin-coated onto the resonator chip before a polished YSO crystal is pressed and clamped on top. The simulated fundamental TE mode demonstrates the evanescent coupling between the cavity mode and erbium-doped YSO.
    \textbf{(c)} Cavity resonances observed in the source (top) and memory (bottom) resonators. The memory resonator spectrum is taken at a cryogenic temperature, and the source resonance spectrum is thermally tuned to match the signal and memory modes.
    \textbf{(d)} Source signal resonance (solid purple) and memory resonance (solid orange) coupled to erbium ensembles. The source resonance was fitted to obtain the measured loaded $Q$ factor of \num{1.32e6}, and the coupled memory resonance was fitted (dotted black) to obtain a cavity-ensemble coupling cooperativity of 1.9.
    \textbf{(e)} Schematic of edge coupling to the resonator chip, allowing blind alignment within a dilution refrigerator. Also shown is a scanning electron microscope image of a SiC microring resonator; the scale bar represents 100 \si{\micro\m}.
    }
    \label{fig:system_diagram}
\end{figure*}

Our integrated source-memory platform is illustrated in Fig.~\ref{fig:system_diagram}a. The architecture consists of two SiC microring resonators fabricated on the same 4H-SiC-on-insulator wafer using an identical lithographic design and fabrication process~\cite{lukin20204h}. Silicon carbide combines low-loss, high-quality-factor resonators, efficient Kerr nonlinearity, and compatibility with heterogeneous rare-earth integration, making it well suited for simultaneously realizing nonlinear photon-pair generation and cavity-enhanced quantum memories. By constructing both subsystems from the same photonic building block, the source and memory inherit nearly identical optical properties, providing intrinsic spectral compatibility for integrated quantum devices.

The operation of the device is shown schematically in Fig.~\ref{fig:system_diagram}a. In the source resonator, a continuous-wave or pulsed pump drives SFWM to generate non-degenerate signal and idler photons~\cite{clemmen2009continuous,zhang2017monolithic,lukin20204h,rahmouni2024entangled}. The signal photon is generated on a resonance that is spectrally aligned to the memory resonator and subsequently routed to the memory chip, while the idler photon is separated using dense wavelength-division multiplexing (DWDM) filters and directed to a single-photon detector. The memory resonator is coupled to an ensemble of \ch{^{167}Er^{3+}} ions through evanescent coupling with the resonator mode, as illustrated in the inset of Fig.~\ref{fig:system_diagram}b. This coupling is realized by directly bonding a polished \eryso crystal onto the silicon-carbide device planarized by a thin hydrogen silsesquioxane (HSQ) layer (Methods), enabling strong couplings between the cavity field and the rare-earth ensemble while preserving the low-loss integrated photonic geometry~\cite{zhong2017nanophotonic,Yang2021,craiciu2019nanophotonic,duranti2024efficient}.

The two resonators are interconnected through a fiber optical link. Light is coupled on- and off-chip using a matched aspheric lens pair in a 1:1 imaging configuration (Methods), enabling efficient transfer between the source and memory subsystems and ease of in-situ alignment in a dilution refrigerator. Following photon-pair generation, residual pump light is rejected using cascaded DWDM filters, while the signal and idler channels are independently routed for memory storage and correlation measurements.

A central feature of the architecture is the spectral matching between source and memory. Resonances of the two microrings are first aligned through temperature tuning of the source resonator, while the erbium ensemble is subsequently tuned into resonance with the memory cavity by applying a magnetic field at millikelvin temperatures. Figures~\ref{fig:system_diagram}c and \ref{fig:system_diagram}d show the resulting spectral alignment between the source resonance, memory resonance, and erbium absorption feature. Because both resonators originate from the same device design, only modest tuning is required to achieve simultaneous alignment of all three subsystems. The signal resonance was selected such that its room-temperature cavity frequency approximately matched the optical $Z_1 \leftrightarrow Y_1$ transition of erbium ions occupying site 2 in \ch{Y2SiO5} at 1539 nm (194.8~THz).

The dual ring architecture was specifically designed to address the bandwidth mismatch that commonly limits source-memory interfaces~\cite{lago2021telecom,liu2021heralded,rakonjac2023transmission}. All microring resonators were fabricated from 2.5~\si{\micro\m}-wide rectangular waveguides and had a diameter of 220~\si{\micro\m}. The source resonance spectrum is shown in Fig.~\ref{fig:system_diagram}c, where we identify high-$Q$ pump, signal, and idler resonances separated by one free spectral range of approximately 160 GHz. The loaded quality factors of the signal, pump, and idler modes are \num{1.32e6}, \num{1.24e6}, and \num{1.23e6}, respectively, corresponding to photon linewidths of 147-158~MHz. These narrow resonances simultaneously enhance SFWM and define the spectral properties of the generated photon pairs. In contrast, bonding of the erbium crystal intentionally increases the loss of the memory resonator, broadening its linewidth while enabling strong collective coupling between the cavity field and the rare-earth ensemble. The resulting memory bandwidth exceeds the photon bandwidth, ensuring efficient capture of the generated signal photons without spectral filtering. From measurements of the cavity linewidth and ensemble coupling strength after hyperfine spin initialization (Supplementary Information S3.1), we extract an ensemble cooperativity of 1.9 over a bandwidth of $\approx$200~MHz.

\begin{figure*}[]
    \centering
    \includegraphics{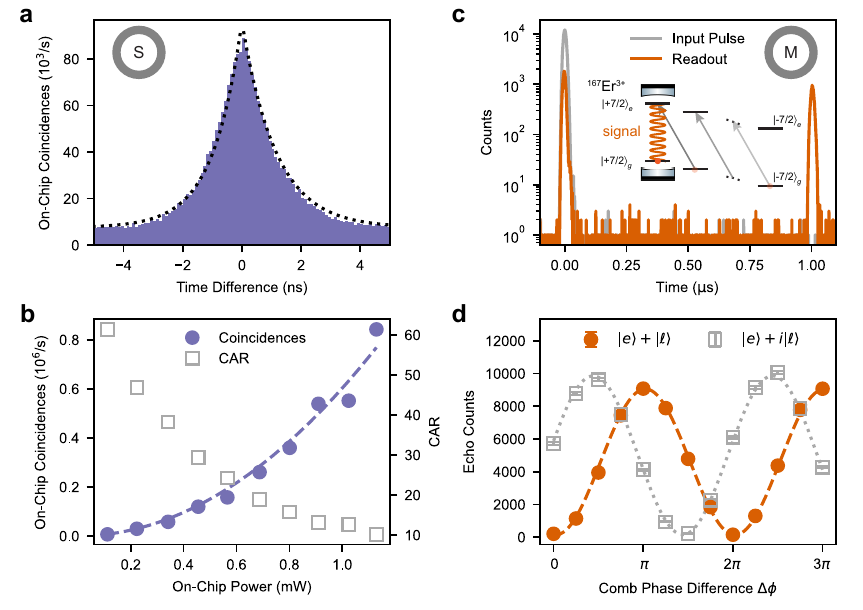}
    \caption{\textbf{Photonic entanglement generation and qubit storage.} \textbf{(a)} Coincidence histogram for signal-idler photon pairs from the source resonator (bars) with fit (dotted line) for 100 ps time bins and an on-chip pump power of 1.1 mW. The two-photon correlation time is 2.1 ns. \textbf{(b)} Measured coincidence count rate (circles) with quadratic fit (dashed line) and Coincidence-to-Accidental Ratio (CAR) (squares) versus on-chip pump power for 1 ns time bins. The fit reveals a pair rate of 6.0$\times10^{5}$~\si{\per\s\per\milli\W\squared} on-chip pump power. \textbf{(c)} Storage and retrieval of weak coherent pulses from the AFC memory resonator with an overall efficiency of 7.7\%. Inset shows \ch{^{167}Er^{3+}} hyperfine initialization process, where a laser drives the $\Delta m = +1$ transitions to prepare the ensemble in the $\ket{+7/2}$ hyperfine spin state on resonance with the memory resonator. \textbf{(d)} Interference of time-bin echoes using two different combs with storage time difference $1/\Delta_1 - 1/\Delta_2$ = 100 ns. The interference for the $|e\rangle + |\ell\rangle$ (circles) and the $|e\rangle + i|\ell\rangle$ (squares) time-bin qubit states are plotted while sweeping the comb phase difference $\Delta\phi$. The two curves were fit to a sinusoid (dashed and dotted lines) with a visibility of 96.9 $\pm$ 2\%.}
    \label{fig:device_characterize}
\end{figure*}

We first characterize photon-pair generation in the source resonator. A continuous-wave (CW) laser at the pump resonance drives SFWM, generating signal and idler photons that are separated using DWDM filters and detected with superconducting nanowire single-photon detectors (SNSPD). A coincidence histogram acquired at an on-chip pump power of 1.1~mW is shown in Fig.~\ref{fig:device_characterize}a. Fitting the coincidence peak yields a two-photon coherence time of 2.1~ns, confirming the narrow bandwidth of the entangled photons. Figure~\ref{fig:device_characterize}b summarizes the source resonator performance. We measure an on-chip pair generation rate of 6.0$\times10^5$~\si{\per\s\per\milli\W\squared} together with a high coincidence-to-accidental ratio of 60 at 112~\si{\micro\W} on-chip pump power. Two-photon Franson interference measurement~\cite{Franson1989} was performed under a low power CW pump (mean pair number 0.013), yielding an accidental-subtracted visibility 96.2 $\pm$ 3.5\% (Supplementary Information S2.2) and confirming high quality of the photonic entanglement. The resulting pair-generation capability is comparable to that of state-of-the-art integrated SiC photon-pair sources~\cite{rahmouni2024entangled}, while simultaneously providing intrinsic spectral compatibility with the cavity-enhanced memory based on the same photonic platform.

Next, we demonstrate cavity-enhanced photonic storage with the memory resonator. Our quantum memory is based on an atomic frequency comb (AFC) in \eryso with a 50 ppm dopant concentration, a well-established platform for optical quantum memories~\cite{jiang2023quantum, craiciu2019nanophotonic, craiciu2021multifunctional}. The \eryso crystal was bonded to the SiC resonator chip using a thin HSQ adhesive layer, which ensures optimal evanescent coupling between the erbium dopants and the resonators (Methods). The bonded device was secured with a clamping mechanism (Supplementary Information S1.1) and was loaded into a dilution refrigerator with a base temperature of 7~\si{\milli\kelvin} and exposed to a magnetic field of 1.238~\si{\tesla} along the crystal $D_1$ axis (see Supplementary Figure~\ref{fig:exp_setup_full} for a full description of the setup) for precise resonance matching between the \ch{^{167}Er^{3+}} optical transition and the memory resonator.

The resonator plays a dual role by enhancing the interaction between the optical field and the Er ensemble while simultaneously defining the memory bandwidth. Following optical pumping and efficient spin initialization into the $|m=+7/2\rangle$ hyperfine state (Supplementary Information S3.1), we characterize the cavity-ion coupling through measurements of the cavity transmission spectrum (Fig.~\ref{fig:system_diagram}d). Fitting the spectrum yields a collective coupling strength of 266 $\pm$ 1~MHz and a cooperativity of 1.87 $\pm$ 0.03 (Supplementary Information S3.1), a 3.4-fold increase over that without hyperfine initialization and placing the device above the unit-cooperativity threshold for efficient cavity-enhanced AFC storage~\cite{Afzelius2010, Moiseev2010}. AFCs are prepared using phase-modulated optical pumping to generate programmable square-teeth comb structures (fixed finesse of 2) with variable bandwidth and storage time~\cite{businger2022non} (Methods). All optical pumping, initialization, and storage operations are performed through the waveguide coupled to the resonator.

Operating as a standalone memory, we assess the memory resonator using weak coherent pulses with a mean photon number of 0.2 and a pulse duration of 15~ns. Total AFC storage-and-retrieval efficiency of 7.7\% was measured at a storage time of $1/\Delta =$ 1~\si{\micro\s} with a 100~MHz comb bandwidth, as shown in Fig.~\ref{fig:device_characterize}c. In comparison, the efficiency was 2.0\% without the hyperfine initialization. Storage times as high as 10~\si{\micro\s} were also achieved (Supplementary Information S3.3, Fig.~S8). More importantly, taking into account the 58\% resonator over-coupling to the waveguide and intrinsic AFC dephasing, we estimate that 89\% of the intra-cavity photons are transferred to the AFC excitation (Supplementary Information S3.3), highlighting an enhanced storage process by the strong collective cavity-ion coupling.

To evaluate memory coherence, we store and analyze time-bin qubits encoded in weak coherent states. Early ($\ket{e}$) and late ($\ket{\ell}$) temporal modes separated by $\Delta\tau$ = 100~ns are prepared using acousto-optic modulators, while a phase modulator controls the relative phase between the two time bins. Interference is performed using the dual-comb AFC protocol~\cite{de2008solid}, in which two AFCs with storage times of 1000~ns and 900~ns are simultaneously prepared such that their retrieval-time difference matches the time-bin separation. As shown in Fig.~\ref{fig:device_characterize}d, we observe an interference visibility of 96.9 $\pm$ 2.0\% for both ($\ket{e}+\ket{\ell}$) and ($\ket{e}+i\ket{\ell}$) input states, confirming faithful preservation of qubit coherence during storage. To further quantify single-photon storage fidelity, we perform a decoy-state analysis following Refs.~\cite{Sinclair2014,zhong2017nanophotonic}. Using three (two plus a vacuum) input photon-number distributions for each qubit state, we derive a lower bound on the single-photon storage fidelity exceeding 98.5\% (Supplementary Information S3.5). This result approaches that of the best bulk AFC memories~\cite{duranti2024efficient,meng2026efficient}, demonstrating that the integrated resonator architecture is fully capable of the high fidelity quantum storage needed to support photon-memory entanglement.

\begin{figure*}
    \centering
    \includegraphics[width=\textwidth]{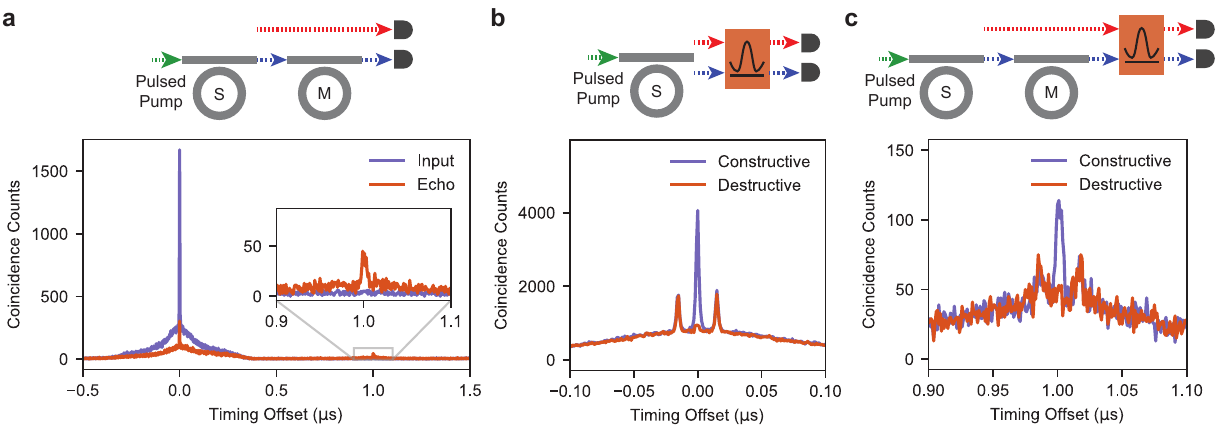}
    \caption{\textbf{Entanglement retrieval and verification.} \textbf{(a)} AFC storage and retrieval of entangled photons with a pulsed pump. After photon pair generation in the source resonator, the signal photon is stored and retrieved from the memory resonator, giving rise to a delayed coincidence peak at the 1~\si{\micro\s} AFC storage time. \textbf{(b)} Two-photon Franson interference of time-energy entanglement generated by the source resonator with a pulsed pump and a mean pair number of 0.24, showing single-pair visibility of 89.6 $\pm$ 2.0\%. \textbf{(c)} Franson interference of the idler and memory-retrieved signal photons reveals a single-pair visibility of 88.1 $\pm$ 8.2\% for the same pump in \textbf{(b)}, demonstrating preservation of the generated entanglement.
    }
    \label{fig:ent_storage}
\end{figure*}

\begin{figure}
    \centering
    \includegraphics[width=\linewidth]{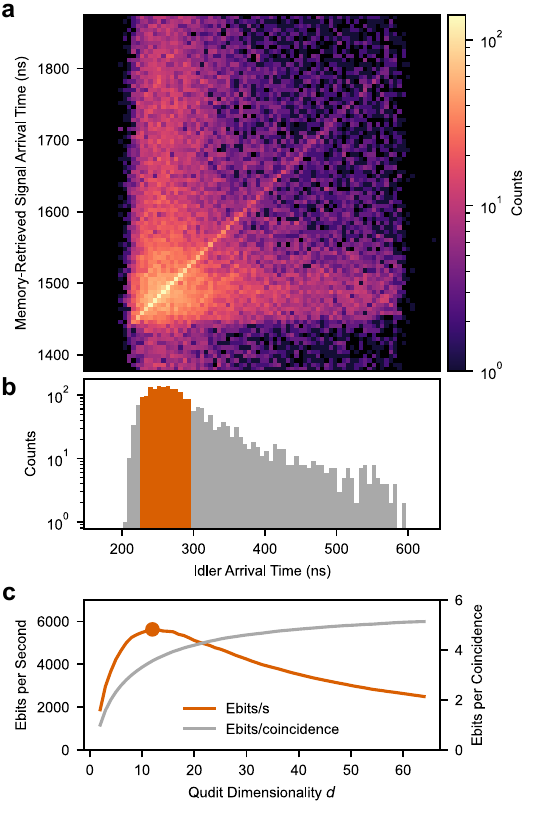}
    \caption{\textbf{Qudit entanglement generation.} \textbf{(a)} Joint temporal correlation between the idler photon and the memory-retrieved signal photon. Each time bin is 6 ns. \textbf{(b)} Coincidence histogram along the diagonal of \textbf{(a)}. The orange-highlighted bins form a qudit frame with $d =$ 12, corresponding to the qudit state achieving a maximal rate in \textbf{(c)}. \textbf{(c)} Photon-memory qudit entanglement generation rates (orange curve) and the photon information efficiency (gray curve) with varying qudit frame dimensionalities. The maximum rate is for $d =$~12 (circle).}
    \label{fig:qudit}
\end{figure}

Having established a spectrally matched source-memory interface, we next demonstrate the central functionality required for absorptive quantum repeaters: generation of entanglement between a telecom photon and a stationary quantum memory. The source resonator is first temperature tuned such that the signal resonance is aligned to the center frequency of the AFC memory. To accommodate the 147~MHz signal photon linewidth, the AFC bandwidth is increased to 200~MHz. Photon pairs are generated using a pulsed pump synchronized to the AFC storage cycle, a crucial requirement to ensure high signal-to-noise ratio at photon retrieval, with the maximal pump pulse duration being half of the storage time~\cite{Hanni2025}. In practice, the effective pump duration was shorter, limited by the rise-time of the modulator and thermal instability of the source resonator. The pump therefore exhibits a non-uniform temporal envelope, as reflected in the arrival-time distribution of Fig.~\ref{fig:qudit}b, which spans a 378~ns window used for the qudit analysis later and lies within the 500~ns set by half the AFC storage time. The signal photon is mapped onto a collective excitation of the erbium ensemble while the idler photon remains in the optical mode, generating time-energy entanglement between a propagating telecom photon and a solid-state quantum memory. Following AFC retrieval, coincidences between the idler photon and the retrieved signal photon are recorded in Fig.~\ref{fig:ent_storage}a. We measure a cross-correlation of $g^{(2)}(1/\Delta)=$ 3.6 $\pm$ 0.3 for the AFC echo, exceeding the Cauchy-Schwarz bound for classical fields, confirming preservation of non-classical correlations in the storage process. We note that multi-pair events contribute significantly to the accidental coincidences given a relatively high source mean pair number $\lambda_\mathrm{p}$ of 0.24 (Supplementary Information S2.1). Pumping the source at a lower $\lambda_\mathrm{p}$ raises the cross-correlation at the expense of rate. The value here reflects our deliberate choice to maximize entanglement throughput.

We next verify photon-memory entanglement by characterizing the underlying continuous-variable (CV) time-energy entanglement of the idler and memory-retrieved signal photons using Franson interferometry~\cite{Franson1989}. Both photons are sent through a shared unbalanced Mach-Zehnder interferometer with a path length difference of $\Delta t =$ 15.3~ns, and the resulting coincidence histograms are shown in Fig.~\ref{fig:ent_storage}b, c. The two outer peaks correspond to distinguishable events, whereas the central peak arises from the indistinguishable short-short and long-long pathways and therefore depends on the interferometer phase, giving rise to two-photon interference. This entanglement resides in the conjugate time and frequency degrees of freedom of the photon pair, and we quantify it through the Franson visibility, using the established relation between fringe contrast and the time-frequency covariance of the pair~\cite{Zhang2014,Zhong2015}. Because that relation holds only for the single-pair component, we isolate it by subtracting the multi-pair accidental background, a physically distinct contribution characterized independently of the interference measurement (Supplementary Information S2.1, Methods). The resulting single-pair visibility thus quantifies the intrinsic quality of the underlying time-energy entanglement, rather than providing a loophole-free certification of it.

To establish a reference visibility, we first perform the interference measurement without memory storage. Locking the interferometer phase using the pump laser as a reference yields constructive and destructive two-photon interference fringes shown in Fig.~\ref{fig:ent_storage}b. The pulsed nature of the pump and a high mean pair number resulted in a multi-pair dominated, triangular accidental coincidence background. To benchmark the intrinsic quality of the time-energy entanglement, we subtracted this accidental coincidence background (Methods), extracting a single-pair visibility of 89.6 $\pm$ 2.0\%. This lower visibility compared with CW pumping is consistent with the finite pump pulse duration, which reduces the two-photon coherence available at the interferometer delay~\cite{Franson1989, Zhong2015}. We then implemented the AFC storage for the signal photon, causing all coincidence peaks to be delayed by the storage time. The resulting interference data are shown in Fig.~\ref{fig:ent_storage}c. We obtain a single-pair visibility of 88.1 $\pm$ 8.2\% after subtraction of accidental coincidences. The consistent visibilities before and after storage evidence faithful preservation of entanglement between all the temporal mode pairs within the pump pulse duration, establishing a quantum interface between the integrated pair source and the quantum memory.

The platform naturally enables high-dimensional photon-memory entanglement by combining the intrinsic temporal multimodality of the AFC memory with the strong temporal correlations of the photon pairs. High-dimensional temporal encodings provide a powerful resource for quantum networking, enabling increased information capacity, improved noise tolerance, and enhanced photon information efficiency compared to qubit-based protocols~\cite{Cozzolino2019,Erhard2020}. In the low-photon-flux regime relevant to long-distance quantum transmission, such encodings can substantially increase the number of entanglement bits (Ebits) delivered per detected photon~\cite{Zhong2015}.

To demonstrate this capability, we record the joint temporal correlations between the idler and memory-retrieved signal photons in Fig.~\ref{fig:qudit}a, revealing a strong temporal correlation along the diagonal, as plotted in Fig.~\ref{fig:qudit}b. We choose a coincidence window of 6~ns to minimize timing error and detector jitters (Supplementary Information S4). We then partition the coincidence histogram in Fig.~\ref{fig:qudit}b into temporal frames containing between $d$ = 2 to 63 time bins and evaluate both the Ebits per photon and the corresponding entanglement rates. The high dimensional entangled qudit state is of the form $|\psi\rangle = \sum_{n=1}^d a_n|n\rangle_s|n\rangle_i$, for which the photon information efficiency is quantified by the Shannon entropy
$H = -\sum_{n=1}^{d} p_n \log_2 p_n$, where $p_n = |a_n|^2$ is the normalized coincidence probability in time bin $n$ of the histogram in Fig.~\ref{fig:qudit}b.

The conjugate variables of CV time-energy entanglement are the arrival-time difference and the frequency sum of the pair~\cite{Zhang2014}. Figure~\ref{fig:qudit}a probes the former and the Franson interference of Fig.~\ref{fig:ent_storage}c checks the latter, so that the two constrain complementary quadratures of the time-frequency covariance matrix~\cite{Zhong2015}. Because the interferometer delay $\Delta t = 15.3$~ns exceeds the 6~ns bin, the Franson interference demonstrates coherence between distinct temporal modes. Within the Gaussian description of the biphoton, which replaces the mutually-unbiased-basis requirement of discrete qudits, these measurements bound the entanglement and its dimensionality. The accessible $d \leq 63$ is limited by the binning rather than by the entanglement itself, whose time-bandwidth product implies a considerably larger Schmidt number, so our reported photon information efficiency is conservative.

We define the on-chip photon-memory entanglement rate based on the rate at which an idler photon is generated in the source waveguide while the corresponding signal echo photon is retrieved into the memory waveguide. This treats the source chip, fiber link and memory chip as one module, so that losses internal to the module -- in particular the fiber-to-chip couplings at both facets -- are retained, while the idler and memory out-coupling and detection paths are divided out using the independently characterized efficiencies in Supplementary Table S1. For each qudit dimensionality $d$, the frame window is scanned across the measured coincidence distribution to maximize the rate. The results are shown in Fig.~\ref{fig:qudit}c. By varying the qudit dimensionality, the platform can generate qudit entanglement optimized for different operating regimes. Using the full temporal extent of the pump pulse ($d$ = 63), we obtain a maximum photon information efficiency of 5.1 Ebits per detected photon, which is advantageous in photon-starved long-distance links where each transmitted photon is a valuable resource. Alternatively, a frame with $d$ = 12, highlighted in Figure~\ref{fig:qudit}b, maximizes the on-chip entanglement generation rate, yielding 5.6~kEbits~\si{\per\second}. These results demonstrate that the same integrated source-memory platform can flexibly generate high-dimensional photon-memory entanglement optimized either for maximum information capacity or peak entanglement throughput, providing a versatile resource for memory-enabled quantum networks.

\section*{Discussion}

In this work, we have demonstrated an integrated photonic architecture capable of generating photon-memory entanglement based on dual micro-ring resonators sharing a common platform. In contrast to previous implementations based on discrete photon sources and quantum memories platforms~\cite{lago2021telecom,liu2021heralded,rakonjac2023transmission}, the dual ring resonator design builds spectral compatibility directly into the hardware, eliminating the need for spectral filtering, bandwidth modification or frequency conversion. The resulting system combines storage of entangled photons, high-visibility photon-memory entanglement, and high-dimensional temporal encoding with photon information efficiencies exceeding five Ebits per detected photon. These capabilities enable a practical integrated source-memory interface for quantum repeaters and provide a scalable foundation for future quantum networking hardware.

Compared on a common external-loss-corrected basis, our entanglement generation rate is among the highest reported for absorptive photon-memory entanglement systems~\cite{lago2021telecom,liu2021heralded,saglamyurek2016multiplexed}. Notably, this rate is presently limited by the inter-chip coupling losses (Supplementary Information S1), rather than by the intrinsic performance of either the photon-pair source or the quantum memory. Consequently, improvements in fiber-waveguide coupling to 50-70\% translate directly into two orders-of-magnitude (one order per coupling) higher entanglement rates, potentially reaching the MEbit~\si{\per\s} regime and enabling operation at a low mean pair number favored by quantum repeater protocols~\cite{simon2007quantum}. Beyond improved coupling efficiency, the integrated architecture naturally extends to a quantum link, where interference between photons emitted from two independent source-memory modules generates remote entanglement between spatially separated memories. The efficient $^{167}$Er hyperfine initialization demonstrated through the waveguide-cavity interface fulfills a prerequisite for future implementation of AFC spin-wave storage~\cite{Afzelius2009,Rani2018} required by on-demand retrieval in a quantum repeater. Looking ahead, the strong source-memory compatibility demonstrated here motivates a complete integration of photon generation and storage within a single resonator, enabling in-situ photon-memory entanglement generation without transmission loss between separate sub-systems ~\cite{Lau2025insitu}. While demonstrated in SiC, the architectural principles are not limited to this material platform. High-quality-factor microring resonators supporting efficient photon-pair generation have been realized in a wide range of integrated nonlinear photonic platforms, including lithium niobate~\cite{zhang2017monolithic}, silicon~\cite{clemmen2009continuous}, silicon nitride~\cite{jiang2023quantum}, and indium gallium phosphide~\cite{akin2024ingap}. Together with the material-agnostic heterogeneous integration approach demonstrated here, which enables evanescent coupling between rare-earth crystals and integrated photonic circuits, these platforms establish a general framework for realizing memory-enabled integrated quantum photonic systems with enhanced performance and functionality. More broadly, the combination of cavity-QED light-matter interface and integrated quantum light generation establishes a versatile platform for studying fundamental light-matter interactions, including quantum-enhanced spectroscopy of many-body atomic systems via quantum illumination~\cite{Lloyd2008}, where the signal photon interrogates the atomic ensembles while the non-interacting idler is to be recombined with the returned signal photon (e.g. echo) to extract correlations arising exclusively from the atomic states. Further, the dual-ring architecture facilitates the realization of squeezing-enhanced light-matter coupling \cite{qin2018exponentially, leroux2018enhancing} by having one resonator parametrically coupled to the matter system while the other provides squeezed reservoir~\cite{Trung25}. Together, these opportunities position the dual-ring platform as a promising foundation for both integrated telecom quantum networks and explorations of quantum many-body photonics.

\section*{Methods}

\subsection*{Device fabrication, bonding and assembly}

Devices were fabricated in a 4H silicon-carbide-on-insulator platform~\cite{lukin20204h}. Bulk 4H-SiC was bonded to an oxidized silicon handle wafer and subsequently thinned to form a 300 nm device layer. Microring resonators and waveguides were patterned by electron-beam lithography using hydrogen silsesquioxane (HSQ) resist and transferred into the SiC layer using \ch{SF6}/\ch{O2} reactive-ion etching. To facilitate removal of the HSQ mask after fabrication, a thin polymethyl methacrylate (PMMA) release layer was introduced between the HSQ and the SiC surface. The PMMA was subsequently dissolved in solvent, enabling lift-off of the HSQ mask.

For memory devices, a thin HSQ spacer layer with a target thickness of approximately 50 nm was spin-coated onto the fabricated resonator chip. A \eryso crystal (Teledyne FLIR) was polished to a root-mean-square (RMS) surface roughness $<$5~\si{\angstrom} and bonded onto the device such that the crystal $b$-axis was oriented perpendicular to the chip surface. The assembly was further secured using a mechanical clamp consisting of screws and an intermediate PTFE cushion layer, which reduced strain on both the crystal and the photonic device (Supplementary Information S1.1).

\subsection*{Cryogenic optical coupling}
Efficient optical access to the integrated devices inside the dilution refrigerator was achieved using an imaging-based coupling scheme. Light was delivered to and collected from the chip using lensed fibers with a focused spot diameter of 2~\si{\micro\meter}. Rather than positioning the lensed fibers directly adjacent to the chip facet, the fiber mode was relayed onto the edge terminated waveguides using a pair of matched high numerical-aperture (NA) aspheric lenses (Thorlabs C330TMD-C) arranged in a 1:1 imaging configuration.

This approach increased the effective working distance from approximately 11~\si{\micro\meter} to 1.8~\si{\milli\meter}, substantially simplifying alignment within the confined space of the dilution refrigerator while eliminating the risk of contact between the lensed fibers and the chip (Supplementary Information S1.1). Device alignment was performed \textit{in situ} using a stack of three Attocube piezo-actuated nanopositioning stages.

\subsection*{Photon-pair generation and characterization}
Photon pairs were generated through spontaneous four-wave mixing (SFWM) in a silicon carbide microring resonator without \ch{Y2SiO5} bonding. The resonator was mounted on a thermoelectric cooler (TEC) for thermal tuning of the cavity resonances. Pump light from an external-cavity diode laser (Toptica DL Pro) was amplified by an erbium-doped fiber amplifier (EDFA), modulated with an acousto-optic modulator (AOM), and spectrally cleaned using three cascaded 100 GHz dense wavelength-division multiplexing (DWDM) filters. For the photon-memory entanglement measurements the AOM gated the pump into pulses synchronized to the AFC storage cycle, at a repetition period of 2~\si{\micro\s} (500~kHz). A 99:1 fiber beamsplitter monitored the input pump power. After the resonator, signal and idler photons were separated and filtered using two sets of four cascaded 200 GHz DWDM filters before detection with superconducting nanowire single-photon detectors (SNSPDs). Photon arrival times were recorded using a time tagger (Swabian Instruments). The pump rejection ports of the DWDM filters were used to monitor the transmitted pump power and determine the fiber-to-waveguide coupling efficiency.

\subsection*{Quantum memory preparation and characterization}
For quantum memory experiments, bonded resonators were operated in a dilution refrigerator (Bluefors LD-50) with a base temperature of 7 mK and equipped with a superconducting vector magnet (AMI). A magnetic field of 1.238 T was applied along the crystal $D_1$-axis to tune the \ch{Er^{3+}} $Z_1 \leftrightarrow Y_1$ optical transition into resonance with the target cavity mode. An external-cavity diode laser (Toptica DL Pro) was swept over a 1 GHz spectral range and injected into the cavity-coupled waveguide for 1 minute to initialize the \ch{^{167}Er^{3+}} ions into the $|m=+7/2\rangle$ hyperfine state.
Atomic Frequency Combs (AFCs) were prepared using a second external-cavity diode laser (Toptica DL Pro) locked to a reference cavity (Stable Laser Systems) via the Pound-Drever-Hall technique. The laser was detuned by $f_s$ = 500 MHz from the cavity resonance and phase modulated using a fiber electro-optic modulator (EOM) to generate a sideband with a spectral profile complementary to the target AFC. The optical field was injected into the waveguide for 5 minutes to create a persistent absorption comb. Unless otherwise noted, AFCs with finesse 2 and tooth spacing $\Delta$ = 1 MHz were used. A secondary comb with tooth spacing 1.111 MHz was employed for the dual-comb interference measurements in Fig.~\ref{fig:device_characterize}d, while a bandwidth of 200 MHz was used for the photon-memory entanglement experiments in Figs.~\ref{fig:ent_storage} and~\ref{fig:qudit}.
Memory characterization was performed using weak coherent probe pulses generated from the initialization laser using three cascaded 200~MHz acousto-optic modulators (AOMs). Probe pulses had a temporal width of 15 ns, a mean photon number of 0.2, and were repeated at a rate of 100 kHz. For dual-comb interference measurements, pairs of pulses separated by 100 ns were generated and the probe laser was locked near the midpoint of the two AFC resonances to ensure a stable relative phase. The AFC sideband frequency $f_s$ was adjusted to control the interferometer phase, while a fiber EOM was used to prepare the $|e\rangle+i|\ell\rangle$ input state by applying a phase shift to the second pulse.

\subsection*{Extracting single-pair visibility}
The photon-memory interference measurements were performed at a high mean pair number to maximize the entanglement generation rate. Under these conditions, accidental coincidences arising from multi-pair events contribute a background to the measured correlations. To quantify the intrinsic time-energy entanglement quality, we therefore report the single-pair visibility after subtracting the accidental-coincidence background.

For the source-only measurements, the background was estimated using the uncorrelated coincidence distribution arising from coincidences between photons of adjacent pulses. When estimating the background for stored pairs, uncorrelated AFC echo from adjacent pulses was used. In order to reduce the uncertainty in the background level, the coincidence distribution associated with this uncorrelated echo was fit with a triangular function, consistent with the convolution of two temporally localized but uncorrelated photon streams. The fitted background was then subtracted from the correlated echo region for both constructive and destructive interference measurements.

Uncertainties of measured visibilities were estimated by propagating both photon counting statistics and background-fit uncertainties. The photon counting uncertainty was verified to follow Poissonian statistics, while the uncertainty in the subtracted background was obtained from the covariance matrix of the triangular fit. These contributions were propagated through the visibility calculation to obtain the reported uncertainty. We note that the Poissonian noise of the photon counts carries more than 95\% of the total uncertainty. Therefore, we expect the uncertainties in Fig.~\ref{fig:ent_storage}b-c to diminish with increasing measurement times. The current data in Fig.~\ref{fig:ent_storage}c was acquired over 20 hours.

\section*{Acknowledgments}

We acknowledge valuable discussions with Hoi-Kwan Lau, Aashish A. Clerk and Yuzhou Chai. The work at Stanford was supported by the Vannevar Bush Faculty Fellowship from the United States Department of War. The work at Chicago was supported by NSF Quantum Leap Challenge Institute for Hybrid Quantum Architectures and Networks (award 2016136) and NSF CAREER award 1944715.

\section*{Author Contributions}

A.K. and I.C. contributed equally to this work. M.G., D.L., A.K., and M.P. designed the devices. D.L. fabricated the devices. A.K., I.C., C.F., and M.P. performed the experiments and collected the data. A.K. and I.C. analyzed the data. A.K., I.C., and T.Z. wrote the manuscript with inputs from all authors. T.Z. and J.V. conceived the project. T.Z. supervised the work.

\section*{Data Availability}
The data supporting the findings of this study are available at doi.org/10.5281/zenodo.21840709.

\section*{Code Availability}
The code used to compute the qudit-frame entropies and rates is available at doi.org/10.5281/zenodo.21840709.

\section*{Competing Interests}
The authors declare no competing interests.

\bibliography{main}

\clearpage

\input{SI}

\end{document}

%% file: SI.tex
\onecolumngrid
\renewcommand{\tocname}{Supplementary Information}
\renewcommand{\appendixname}{Supplement}
\setcounter{equation}{0}
\setcounter{figure}{0}
\setcounter{table}{0}
\renewcommand{\theequation}{S\arabic{equation}}
\renewcommand{\thefigure}{S\arabic{figure}}
\renewcommand{\thesubsection}{S\arabic{subsection}}
\renewcommand{\theHfigure}{SI\arabic{figure}}
\renewcommand{\thetable}{S\arabic{table}}
\renewcommand{\theHtable}{SI\arabic{table}}

\clearpage
\begin{appendices}

\section*{Supplementary Information}

\subsection{Experimental Setup}
\subsubsection{Schematics of the Complete Experimental Setup}
\label{sup:exp_setup}

\begin{figure}
    \centering
    \includegraphics[]{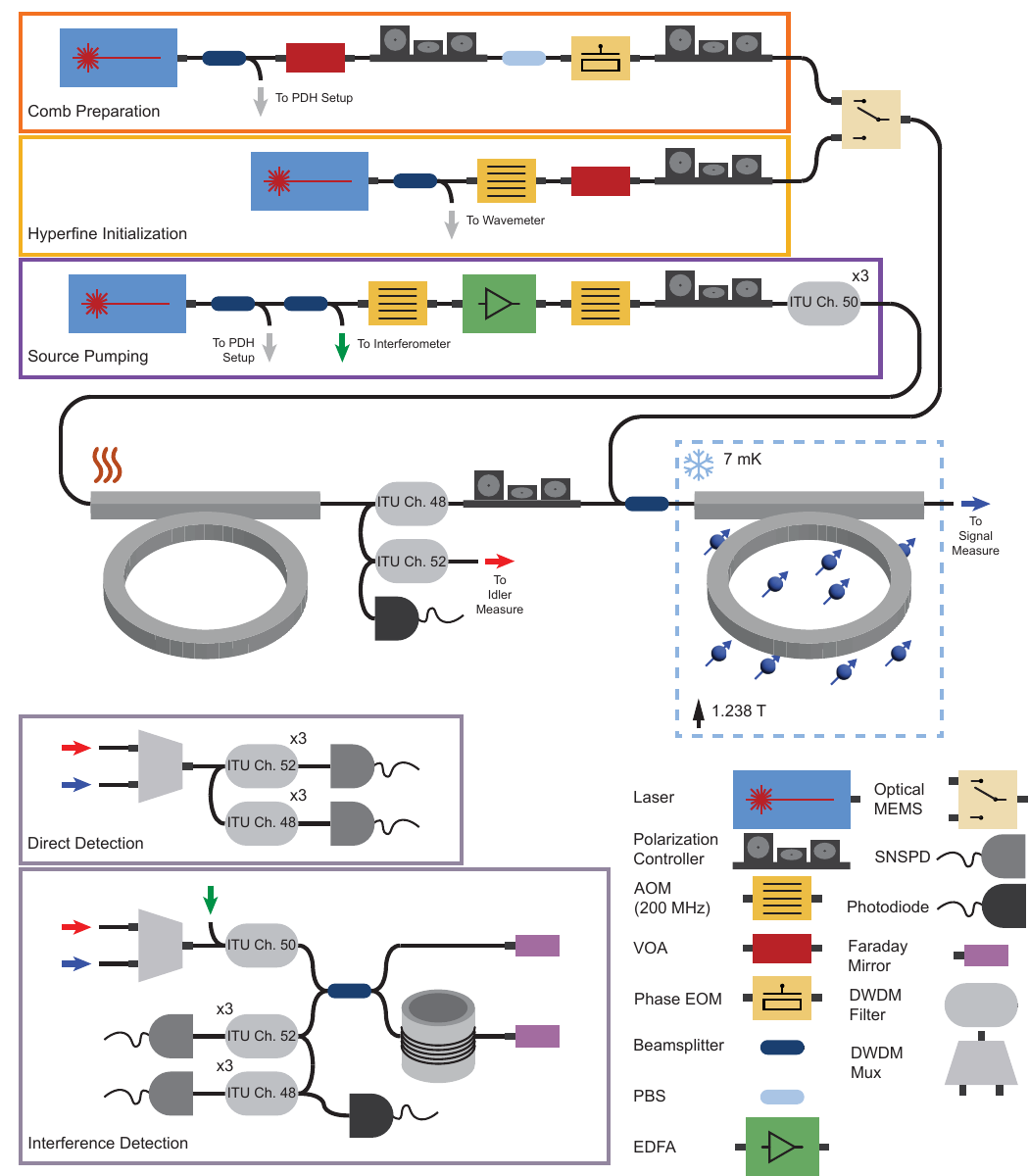}
    \caption{Experimental setup for main text Figures~\ref{fig:ent_storage} and \ref{fig:qudit}. \textbf{Memory Preparation} Hyperfine initialization was performed by sweeping a laser over a 1 GHz frequency range (stabilized using a wavemeter). The AFC was prepared using a second laser locked to a stable reference cavity, with a comb-shaped spectral sideband prepared using a phase EOM. The initialization and comb preparation lasers were switched using an optical MEMS switch, with the AOM and VOA on the initialization path used to suppress output during the main experiment. \textbf{Entangled photon storage} A third frequency-locked laser was used to pump the source resonator, where a 450-ns pulse was carved using cascaded AOMs and amplified using an EDFA. An additional AOM and DWDM filter were used for temporal and spectral filtering. All filters are labeled with their ITU channel numbers. \textbf{Entangled photon Measurement} Signal and idler pairs were separated using cascaded DWDM filters and detected using SNSPDs. For Fig.~\ref{fig:ent_storage}b, c, an unbalanced fiber interferometer was inserted on the shared signal and idler path before detection. The interferometer phase was stabilized by detecting the pump laser phase with a photodiode, and adjusting the relative phase of the interferometer arms with a fiber stretcher. AOM: acoustic-optic modulator; VOA: variable optical attenuator; SNSPD: superconducting nanowire single photon detector; EOM: electro-optic modulator; PBS: polarization beamsplitter; EDFA: Erbium-doped fiber amplifier; DWDM: dense wavelength-division-multiplexing; Mux: multiplexer.}
    \label{fig:exp_setup_full}
\end{figure}

The complete experimental setup is shown in Fig.~\ref{fig:exp_setup_full}. Table~\ref{tab:loss} summarizes the measured efficiencies of the key optical components in the experiment. These values are used throughout the manuscript to calculate on-chip pair generation rates and photon-memory entanglement generation rates from the measured detection statistics.

\subsubsection{Dilution Refrigerator Alignment}
\label{sup:fiber_lens_align}
\begin{figure}
    \centering
    \includegraphics[]{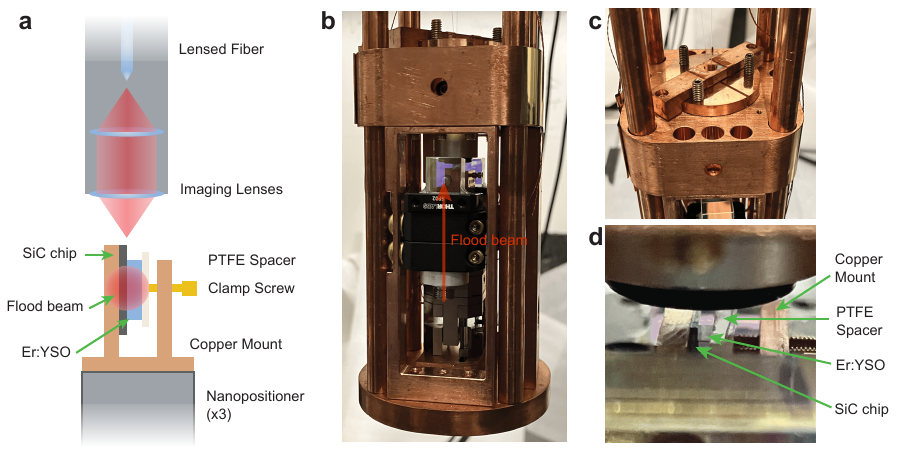}
    \caption{\textbf{Dilution Refrigerator Setup.} \textbf{(a)} Schematic setup for the memory resonator within the 7 mK dilution refrigerator. Two matched aspheric lenses image the lensed fiber pair output to couple to two waveguides on the silicon carbide chip, introducing a $\approx$1 mm working distance to prevent mechanical contact during in-situ alignment. The \eryso crystal is mechanically clamped to the resonator chip using brass clamp screws with an intermediate PTFE cushioning layer. The resonator chip is aligned to the imaged fiber beams using a set of 3 Attocube nanopositioners. A wide flood beam passes through the \ch{Y2SiO5} crystal for monitoring the erbium ensemble during cooling and magnet ramping. \textbf{(b)} Photograph of the fully assembled setup. The silver lens tube and beamsplitter cube shown in the front are used to fiber-couple the flood beam and deliver it to the sample (fiber not shown). The clamping setup is visible behind the beamsplitter cube, and the nanopositioner stack is visible behind the lens tube. \textbf{(c)} Alignment setup for lensed fiber pairs. Three set screws are used to pivot the fiber angle to match the sample waveguide, and a center clamp ensures minimal movement during cooldown. \textbf{(d)} Details of the sample and the clamping mechanism.}
    \label{fig:exp_photos}
\end{figure}

The memory resonator was operated inside a dilution refrigerator at a base temperature of 7 mK. Because direct optical access to the device is unavailable after cooldown, efficient and mechanically stable coupling between optical fibers and the integrated waveguides was achieved through a custom cryogenic alignment assembly, shown in Fig.~\ref{fig:exp_photos}.
Light was coupled onto and off of the chip using a pair of lensed fibers and two high numerical aperture (NA) aspheric lenses arranged in a 1:1 imaging configuration. Both the fiber spacing and the waveguide spacing are 250~\si{\micro\m}. The optical geometry was optimized to maximize coupling efficiency while maintaining sufficient working distance to avoid mechanical contact during alignment. The fiber assembly provided angular adjustment through a set of precision screws, while the sample position was controlled using a stack of three Attocube nanopositioners, enabling alignment along three translational degrees of freedom. The total fiber-to-chip coupling efficiency was $\leq$10\%, limited primarily by aberrations introduced by the imaging optics and by imperfections in the waveguide taper structures. Although these losses reduced the measured photon-memory entanglement rate, they did not affect the intrinsic source and memory performance and represent an engineering limitation that can be addressed through improved photonic packaging and coupling structures.

A secondary free-space “flood beam” propagated through the \eryso crystal perpendicular to the waveguide axis, as shown in Fig.~\ref{fig:exp_photos}a. Although insufficient for coherent control of the ensemble, this beam enabled real-time monitoring of the erbium absorption spectrum during cooldown and magnetic-field ramping. The resulting spectral information provided a convenient method for tuning the magnetic field and aligning the erbium transition with the cavity resonance prior to AFC preparation.
The coupling efficiencies reported in Table~\ref{tab:loss} were measured after cooldown and constitute the dominant loss mechanism in the present experiment. As discussed in the main text, these losses arise primarily from the cryogenic fiber-to-chip interface rather than the intrinsic performance of either the photon source or quantum memory.

\begin{table}[h!]
\centering
\begin{tabular}{|l|l|l|l|}
\hline
\textbf{Loss Element}                & \textbf{Efficiency}  & \textbf{Source Pair Gen. Rate} & \textbf{Photon-Memory Gen. Rate}\\ \hline
Source Waveguide to Fiber, signal & 0.052    & Divided & Retained\\ \hline
Source Waveguide to Fiber, idler  & 0.052    & Divided & Divided\\ \hline
Memory Input Cryostat Feedthrough    & 0.82       & - & Retained\\ \hline
Fiber to Memory Input Waveguide & 0.07      & - & Retained \\ \hline
Memory Output Waveguide to Fiber & 0.07      & - & Divided \\ \hline
Memory Output Cryostat Feedthrough   & 0.85      & - & Divided \\ \hline
Filters + SNSPD Efficiency (Idler)   & 0.33       & Divided & Divided\\ \hline
Filters + SNSPD Efficiency (Signal)  & 0.44       & Divided & Divided\\ \hline
\end{tabular}
\caption{Measured efficiencies for the experimental setup in Figure~\ref{fig:exp_setup_full} and their inclusion in the generation rates calculations.}
\label{tab:loss}
\end{table}

\subsection{Source Resonator Characterization}
\subsubsection{Sources for Accidental Coincidence}
\label{sup:pair_noise}
\begin{figure}
\centering
\includegraphics[]{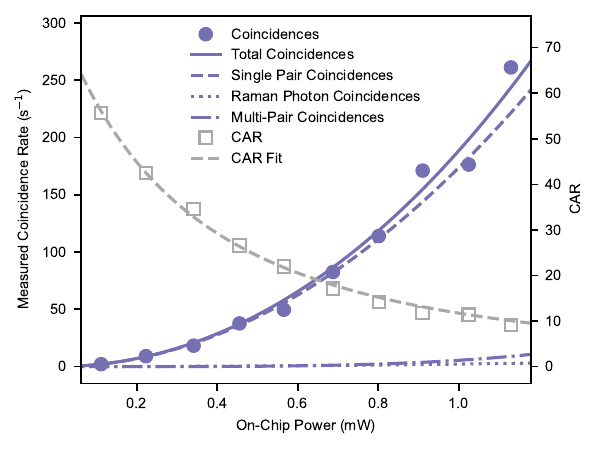}
\caption{\textbf{Noise model fit to the coincidence measurements.} Measured coincidence rates and Coincidence-Accidental Ratio (CAR) together with fits to a model including photon-pair generation, Raman scattering, and detector dark counts. The fitted contributions from different processes are shown separately, revealing that multi-pair events dominate the accidental coincidences under the operating conditions used for photon-memory entanglement generation.}
\label{fig:CAR}
\end{figure}
To identify the dominant sources limiting the CAR and two-photon interference visibility, we model the detected coincidence statistics as arising from three independent processes: spontaneous four-wave-mixing photon-pair generation, Raman scattering, and detector dark counts. Pair generation scales quadratically with pump power, whereas Raman scattering scales linearly. Assuming Poissonian statistics for all processes, the CAR can be expressed as
\begin{equation}
\rm{CAR}=\frac{\lambda_p}{(\lambda_p+\lambda_r+\lambda_d/\eta_s)(\lambda_p+\lambda_r+\lambda_d/\eta_i)},
\label{eq:CAR}
\end{equation}
where $\lambda_p$, $\lambda_r$, and $\lambda_d$ are the mean pair, Raman, and dark-count photon numbers within the coincidence window, and $\eta_s$ and $\eta_i$ are the signal and idler detection efficiencies, respectively.
The measured coincidence rate was first used to determine the pair-generation coefficient, after which the CAR data were fit using Eq.~(\ref{eq:CAR}) to extract the Raman contribution. The resulting fit is shown in Fig.~\ref{fig:CAR}. The model indicates that the dominant accidental contribution originates from multi-pair generation rather than Raman scattering. At the 2~mW pump power used for the photon-memory storage measurements, the extrapolated mean pair number within the 2.1~ns two-photon correlation time is 0.24, and the contribution from multi-pair events far exceeds that of Raman-scattering induced events by nearly an order of magnitude. We therefore conclude that the limited CAR in the storage experiment is primarily limited by multi-pair generation.

\subsubsection{Visibility of the Photon-Pair Source}
\begin{figure}
\centering
\includegraphics[]{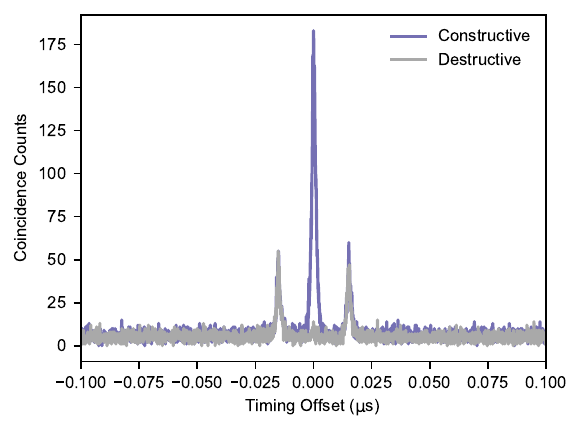}
\caption{\textbf{Franson interference of the photon-pair source.} Coincidence histograms obtained using a Franson interferometer under constructive (blue) and destructive (gray) interference conditions. The three peaks correspond to the distinguishable long--short and short--long path combinations (side peaks) and the interfering short--short and long--long contributions (center peak). From the interference fringe of the center peak, we extract a raw visibility of 90.4 $\pm$ 3.3\% and an accidental-subtracted visibility of 96.2 $\pm$ 3.5\%, for $\lambda_p$ = 0.013.}
\label{fig:source_vis}
\end{figure}

To independently characterize the quality of the time-energy entanglement generated by the source resonator, we performed a Franson interference measurement using the setup shown in Fig.~\ref{fig:exp_setup_full}. The interferometer was implemented as an imbalanced interferometer with Faraday mirrors acting as retroreflectors, suppressing polarization-dependent phase drifts between the two arms. A low-power continuous-wave pump (455 \si{\micro\watt} on-chip) was used to minimize multi-pair generation (mean pair number $\lambda_p$ = 0.013) and isolate the intrinsic entanglement quality of the source.
The resulting coincidence histogram exhibits the characteristic three-peak structure of a Franson interferometer, as shown in Fig.~\ref{fig:source_vis}. The outer peaks correspond to distinguishable long-short and short-long path combinations, while the central peak arises from interference between the indistinguishable long--long and short-short contributions. By varying the interferometer phase, we observe high-contrast interference fringes with a raw visibility of 90.4 $\pm$ 3.3\%, far exceeding the classical threshold and confirming the generation of high quality entangled photon pairs.
After background subtraction, the measured single-pair visibility increases to 96.2 $\pm$ 3.5\%. We attribute the residual degradation to imperfections in the interferometer, including unequal transmission through the two arms arising from the fiber stretcher and associated splices, rather than limitations of the photon-pair source itself.

\subsection{AFC Memory Preparation and Characterization}
\subsubsection{Hyperfine Initialization}
\label{sup:init_memory_holeburning}

Efficient operation of the cavity-enhanced AFC memory requires initialization of the \ch{^{167}Er^{3+}} population into a single hyperfine state. The \eryso optical transition contains eight hyperfine sublevels arising from the nuclear spin of \ch{^{167}Er}, and under thermal equilibrium the population remains distributed among these states. As a result, the optical absorption spectrum is substantially broadened beyond the intrinsic inhomogeneous linewidth, reducing the achievable cavity cooperativity.

\begin{figure}
    \centering
    \includegraphics[]{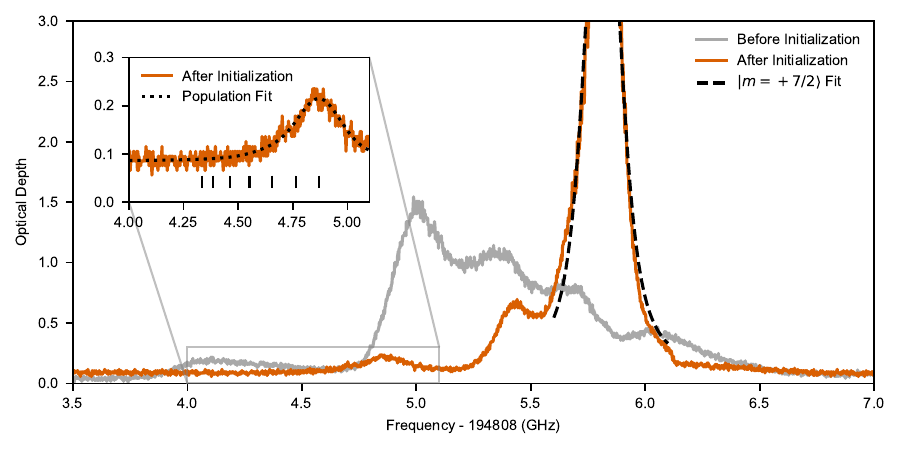}
    \caption{\textbf{Hyperfine initialization of bulk \eryso.}
Optical absorption spectrum of a bulk \ch{^{167}Er^{3+}:Y2SiO5} crystal measured at the same temperature as the main-text experiments and a magnetic field of approximately 300~mT. The gray curve shows the ensemble before initialization, while the orange curve shows the spectrum after optical pumping on the $\Delta m=+1$ transitions. Hyperfine initialization transfers the population into the $|m=+7/2 \rangle$ nuclear-spin state, resulting in a pronounced increase in absorption and a substantial narrowing of the optical spectrum. The dominant  $|m=+7/2\rangle_g \leftrightarrow |m=+7/2\rangle_e$  transition is fit by a Voigt profile (black dashed line), yielding a full-width at half-maximum linewidth of 183~MHz. \textbf{Inset:} Residual absorption on the $\Delta m=-1$ sideband used to estimate the hyperfine-state population. Fitting the sideband spectrum using the hyperfine splittings reported in ~\cite{Rani2018} yields a hyperfine polarization of approximately 98\%, consistent with previous measurements of \ch{^{167}Er^{3+}:Y2SiO5}.
}
    \label{fig:initialize_bulk}
\end{figure}

To characterize the initialization process independently of the cavity, we first performed measurements on a bulk crystal sample cut from the same boule as the device crystal. The measured spectra are shown in Fig.~\ref{fig:initialize_bulk}. Following established techniques~\cite{Rani2018,Stuart2021}, a laser was swept across the $\Delta m = +1$ transitions, optically pumping the ions into the ($m=+7/2$) hyperfine state. Fitting the residual $\Delta m=-1$ sideband yields an estimated hyperfine polarization of approximately 98\%, consistent with previous reports. The resulting optical transition exhibits a linewidth of 183~MHz and an enhancement of the peak optical depth by a factor of 5.7, providing a benchmark for the initialization efficiency achievable in the cavity-coupled device.

Implementing this procedure within a high-$Q$ resonator presents additional challenges as the cavity linewidth is substantially narrower than the hyperfine splitting and the cavity transmission no longer directly reflects the ensemble absorption spectrum. To facilitate initialization, we introduced ``flood beam'' -- a secondary free-space probe beam propagating through the crystal perpendicular to the waveguide axis. This probe provided direct access to the erbium absorption spectrum and enabled alignment of the $m=+7/2$ and $\Delta m=0$ transitions with the cavity resonance through magnetic-field tuning.

\begin{figure}
    \centering
    \includegraphics[]{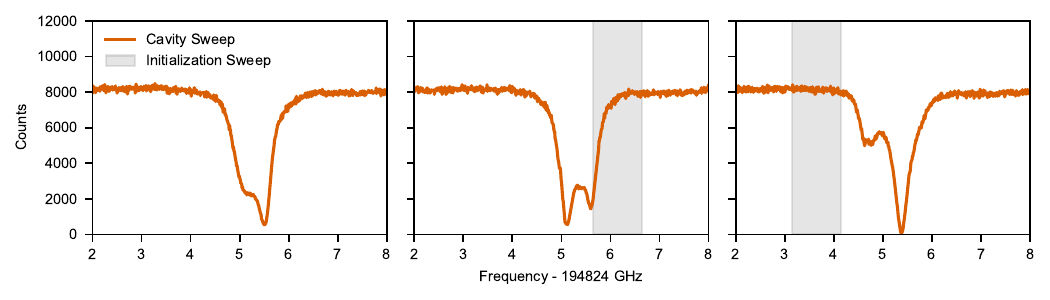}
\caption{\textbf{Hyperfine initialization of the cavity-coupled erbium ensemble.}
Memory cavity transmission spectra measured after optical pumping at different initialization frequencies. The left panel shows the cavity-coupled spectrum before hyperfine initialization, obtained by scanning a weak probe laser across the cavity resonance. The gray shaded regions in the center and right panels indicate the frequency ranges over which the initialization laser was swept. Optical pumping transfers population into different nuclear-spin states, producing distinct cavity-coupled absorption features and demonstrating control over the hyperfine-state distribution of the ensemble. The initialization condition shown in the center panel concentrates population into the m=+7/2 manifold and was used for obtaining Fig.~\ref{fig:system_diagram}d and for all AFC storage experiments reported in the main text.}
    \label{fig:initialize}
\end{figure}

After coarse spectral alignment, hyperfine initialization was performed through the integrated waveguide by sweeping a laser across a 1~GHz frequency range for approximately one minute. The center frequency of the sweep was optimized by monitoring the cavity transmission spectrum using a weak probe field detected with an SNSPD. Representative spectra are shown in Fig.~\ref{fig:initialize}. The chosen operating point produces a narrow ensemble resonance centered within the cavity mode and is used throughout the experiments in the main text.

To quantify the resulting cavity-ensemble coupling, the cavity resonance was first measured with the erbium transition magnetically detuned, yielding a cavity linewidth of $\kappa$ = 396~MHz. The coupled cavity spectrum was then fitted using the ensemble-modified cavity response of Eq.~(\ref{eq:coupling})~\cite{Diniz2011,miyazono2017coupling,Yang2021}, where $\kappa_\mathrm{in}$ is the waveguide coupling, $\omega_0$ the cavity center frequency, $\omega_\mathrm{ion}$ the ensemble center frequency, and the function $W(\omega)$ accounts for the ensemble coupling strength $G$, inhomogeneous linewidth $\Gamma$, and the shape of the inhomogeneous distribution $\rho$. Because the exact ensemble lineshape is unknown and is reportedly a Voigt profile~\cite{Rani2018} (i.e. a convolution of Lorentzian and Gaussian), both Lorentzian and Gaussian inhomogeneous distributions were considered. The extracted parameters are summarized in Table~\ref{tab:coupling}. While both models yield similar collective coupling strengths, the Gaussian fit predicts a narrower ensemble linewidth and correspondingly a larger cooperativity. Throughout the main text we conservatively report the lower Lorentzian value, yielding G = 266 $\pm$ 1~MHz and $\mathcal C$ = 1.87 $\pm$ 0.03.

\begin{equation}
    \label{eq:coupling}
    T(\omega) = \left|1 - \frac{i\kappa_\mathrm{in}}{\omega - \omega_0 + i\kappa/2 - W(\omega - \omega_\mathrm{ion})}\right|^2
\end{equation}

\begin{table}[h]
\centering
\begin{tabular}{|l|l|l|}
\hline
Parameter     & Lorentzian Fit  & Gaussian Fit    \\ \hline
$G$           & 266 $\pm$ 1 MHz & 228 $\pm$ 1 MHz \\ \hline
$\Gamma$      & 384 $\pm$ 5 MHz & 245 $\pm$ 2 MHz \\ \hline
$\mathcal{C}$ & 1.87 $\pm$ 0.03 & 2.15 $\pm$ 0.03 \\ \hline
\end{tabular}
\caption{Fitted values for cavity-ensemble coupling $G$, inhomogeneous linewidth $\Gamma$, and cooperativity $\mathcal{C}$ assuming both a Lorentzian and Gaussian frequency distribution for the ion ensemble.}
\label{tab:coupling}
\end{table}

An independent assessment of the effectiveness of hyperfine initialization can be obtained from the AFC storage efficiency (discussed in Section S3.3). Without initialization, the measured maximum storage efficiency was 2.0\%, corresponding to an estimated comb cooperativity $\mathcal{C}_{\mathrm{comb}}$ of 0.15 based on Eq.~\ref{eq:AFCeff}. With initialization, the storage efficiency increased to 7.7\%, corresponding to a comb cooperativity $\mathcal{C}_{\mathrm{comb}}$ of 0.51, a 3.4-fold enhancement in effective ensemble spectral density arising from the hyperfine pumping. Although this improvement is below the factor of 5.7 from the bulk-crystal measurements, it is sufficient to bring the memory close to the impedance-matched regime, for which $4\mathcal{C}_{\mathrm{comb}}/(\mathcal{C}_{\mathrm{comb}}+1)^2$ = 0.89 versus a value of 1 for perfect impedance matching.

\subsubsection{Programmable AFC Preparation}

The AFC was prepared using the spectral hole-burning technique developed in Ref.~\cite{businger2022non}. Rather than burning individual AFC teeth sequentially, we synthesize a frequency-domain waveform whose spectrum corresponds to the complement of the desired AFC structure. When applied to a phase electro-optic modulator (EOM), this waveform generates a comb-shaped optical sideband that simultaneously pumps ions out of the anti-tooth regions, leaving behind the desired AFC absorption profile.

To generate the anti-tooth spectrum, the RF waveform was constructed as a sum of hyperbolic-secant frequency sweeps,

\begin{equation}
    A_n(t) = \sech(\beta t) \sin\left[ 2\pi(f_0+n\Delta)t+2\pi\frac{\Delta_f}{2\beta}\ln\left(\cosh(\beta t)\right)\right] \qquad t \in [-\tau/2, \tau/2]
\end{equation}
where $n$ denotes the tooth number, $\Delta$ is the AFC tooth spacing, $\Delta_f$ is the anti-tooth width, $f_0$ defines the offset of the comb sideband from the optical carrier, $\tau$ is the pulse repetition period, and $\beta$ is a parameter of the sweep speed. The AFC finesse is controlled through the ratio $\Delta/\Delta_f$. Individual frequency sweeps corresponding to different anti-teeth are temporally offset before being combined, suppressing interference in the time domain while preserving the desired frequency-domain structure.

\begin{figure}
    \centering
    \includegraphics[]{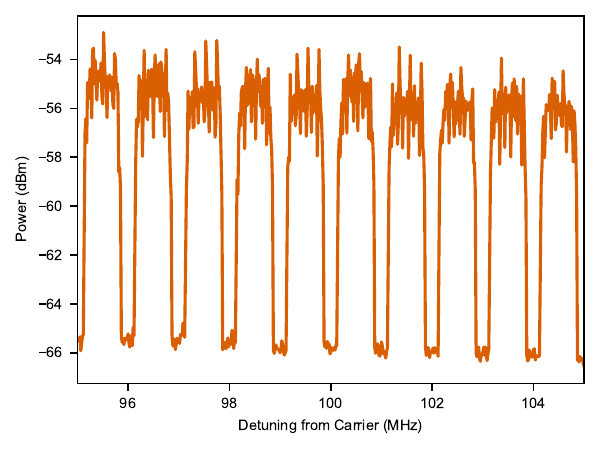}
   \caption{\textbf{RF spectrum used for AFC preparation.} Example RF spectrum applied to the phase EOM to generate the comb-shaped optical sideband used for AFC preparation. The spectrum consists of uniformly spaced square anti-teeth whose spacing and width determine the AFC storage time and finesse, respectively. The spectrum shown corresponds to $\Delta$ = 1~MHz, $f_0$ = 100 MHz, and RF parameters $\tau$ = 2 ms, $\beta$ = 10$/\tau$. A finesse $F$ = 3 is shown so that the anti-tooth width and spacing are visually distinct.}
    \label{fig:eom}
\end{figure}

The resulting waveform was synthesized over a 2~ms interval and repeatedly played from a Zurich Instruments HDAWG to drive a phase EOM. Figure~\ref{fig:eom} shows a representative RF spectrum for a 1~MHz-tooth-spacing AFC. The generated spectrum consists of a series of high-contrast square anti-teeth with uniform spacing and width, enabling precise control over the AFC bandwidth, storage time, and finesse. The optical sideband produced by the EOM was subsequently used to burn all AFCs employed in the experiments in the main text.

\subsubsection{AFC Efficiency Analysis}
\label{sup:mem_efficiency}

The efficiency of a cavity-enhanced AFC memory is expressed as~\cite{Moiseev2010, zhong2017nanophotonic},
\begin{equation}
    \eta_\mathrm{AFC} = \eta_\mathrm{store}^2\eta_\mathrm{deph} =  \left(\frac{\kappa_\mathrm{in}}{\kappa} \frac{4\mathcal{C}_\mathrm{comb}}{(\mathcal{C}_\mathrm{comb} + 1)^2}\right)^2 \eta_\mathrm{deph}
    \label{eq:AFCeff}
\end{equation}
\noindent where $\eta_\mathrm{store}$ describes the efficiency of photon transfer (and retrieval) between the waveguide and the AFC, $\kappa_\mathrm{in}/\kappa$ describes coupling between the waveguide and cavity mode, $\mathcal C_{\mathrm{comb}}$ is the AFC cooperativity, and $\eta_\mathrm{deph}$ accounts for dephasing arising from finite AFC finesse and optical decoherence. For the square-tooth AFCs~\cite{Jobez2016} employed in this work,
\begin{equation}
    \eta_\mathrm{deph} = \eta_F\eta_{T_2} = \mathrm{sinc}^2\left(\frac{\pi}{F}\right)\exp\left(-\frac{4}{\Delta T_2^\mathrm{AFC}}\right)
\end{equation}
where $F$ is the AFC finesse and $T_2^\mathrm{AFC}$ is the optical coherence lifetime of the ions participating in the AFC.

\begin{figure}
    \centering
    \includegraphics[]{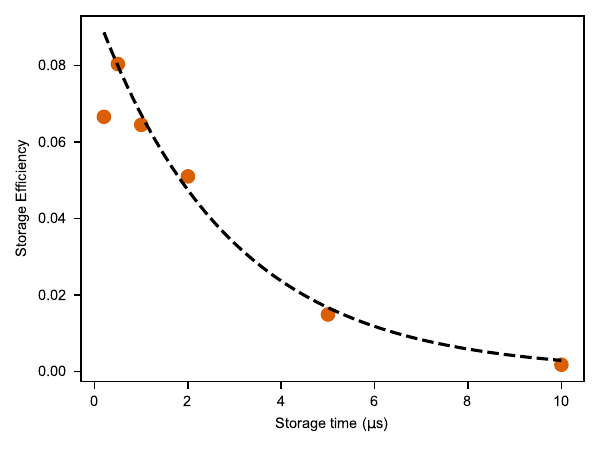}
    \caption{\textbf{Storage efficiency versus AFC storage time.} AFC storage efficiency measured for different storage times while maintaining a constant finesse of $F$=2. The dashed line is a fit to the expected exponential decay arising from optical decoherence, yielding an AFC coherence lifetime of $T_2^\mathrm{AFC}$=11.5~\si{\micro\s}. The shortest-storage-time point is excluded from the fit due to finite input-pulse bandwidth effects.}
    \label{fig:storage_time_sweep}
\end{figure}

To characterize the memory, weak coherent pulses with mean photon number 0.2 were stored and retrieved through the AFC. The storage time was varied while maintaining a constant finesse of $F$ = 2, ensuring that changes in efficiency were dominated by optical decoherence rather than changes in comb structure. The measured storage efficiencies are shown in Fig.~\ref{fig:storage_time_sweep}. Fitting the data yields an AFC coherence lifetime of $T_2^\mathrm{AFC}$ = 11.5~\si{\micro\s}.

The fitted lifetime further yields a zero-storage-time ($\Delta=\infty$) efficiency of $\eta_0=\eta_\mathrm{store}^2\eta_F$ = 10.9\%. Accounting for the finite-finesse dephasing factor $\eta_F = 4/\pi^2$, we infer a storage efficiency of $\eta_\mathrm{store} =$ 0.518, corresponding to an AFC cooperativity of $\mathcal C_\mathrm{comb}\approx$ 0.51. Therefore, we estimate a transfer efficiency of intra-cavity photons to the AFC of 89\%. This value is close to the ideal cavity impedance matching regime, where the transfer efficiency is 1.

\subsubsection{Temporal Multimode Storage}
\label{sup:multimode_storage}

A key advantage of AFC memories is their ability to store multiple temporal modes without sacrificing storage efficiency. To characterize the temporal multimode capacity of the cavity-enhanced memory, we stored a train of ten weak coherent pulses separated by 100 ns.

\begin{figure}
    \centering
    \includegraphics[]{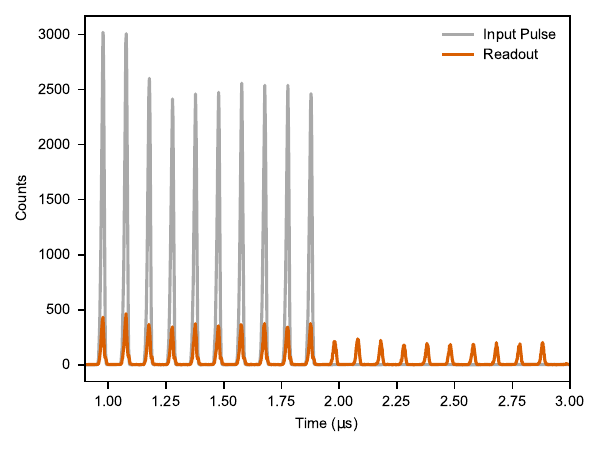}
    \caption{\textbf{Temporal Multimode Storage.} Histogram showing storage of 10 temporal modes in the AFC with 100 ns separation. The average efficiency over all bins is 8.3\%, comparable to the measured storage efficiency of a single input pulse.}
    \label{fig:multimode}
\end{figure}

The retrieved pulse train is shown in Fig.~\ref{fig:multimode}. All ten temporal modes are recovered with an average storage efficiency of 8.3\%, comparable to that obtained for a single stored pulse. The absence of any significant efficiency variation across the pulse train indicates that the AFC stores multiple temporal modes with negligible cross-talk over the duration of the experiment.

While the present demonstration is limited to ten modes by the available pulse-generation electronics, the achievable multimode capacity is fundamentally determined by the ratio of AFC storage time to input pulse duration. The demonstrated temporal multimode operation forms the basis for the high-dimensional photon--memory entanglement and photon information efficiency measurements reported in the main text.

\subsubsection{Decoy-State Analysis of Memory Fidelity}
\label{sup:decoy_state}

To characterize the coherence preservation of the AFC memory, we stored weak coherent time-bin states and measured interference using a dual-comb AFC interferometer~\cite{de2008solid,liu2020demand}. Early and late time bins separated by 100 ns were generated using AOMs, while a phase EOM was used to prepare arbitrary relative phases between the two temporal modes. Interference was realized by preparing two AFCs with storage times differing by 100 ns, causing the retrieved early and late time bins to overlap and interfere upon recall.

\begin{figure}
    \centering
    \includegraphics[]{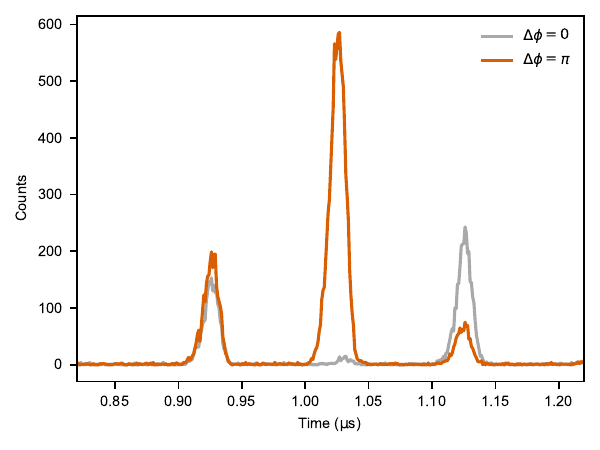}
  \caption{\textbf{Dual-comb AFC interference.} AFC echoes for constructive ($\Delta\phi = 0$) and destructive ($\Delta\phi = \pi$) interference in the AFC interferometer. The strong modulation of the central peak demonstrates preservation of phase coherence during storage and yields an interference visibility of 96.9 $\pm$ 2.0\%.
}
    \label{fig:afc_interference}
\end{figure}

Representative AFC echoes are shown in Fig.~\ref{fig:afc_interference}. By varying the relative phase between the two AFCs, constructive and destructive interference are observed in the central retrieval peak, while the side peaks remain unchanged. From these measurements, we extract an interference visibility of 96.9 $\pm$ 2.0\%, demonstrating excellent preservation of phase coherence during storage.

To estimate the storage fidelity for single-photon qubits, we performed a decoy-state analysis following Ref.~\cite{Sinclair2014}. Weak coherent states with mean photon numbers of 0, 0.12, and 0.27 photons per pulse were stored and analyzed. Applying the decoy-state formalism to the $(|e\rangle+|l\rangle)/\sqrt{2}$ and $(|e\rangle+i|l\rangle)/\sqrt{2}$ input states yields fidelities of 97.1\% and 97.4\%, respectively, corresponding to a lower bound on the single-photon storage fidelity exceeding 98.5\%. These values confirm that the cavity-enhanced AFC preserves quantum coherence at a level comparable to the best bulk rare-earth AFC memories.

\subsection{Qudit Time-bin Optimization}
\label{sup:qudit_error}

To analyze the high-dimensional photon--memory entanglement, the joint temporal correlation of the signal and idler detection times was rebinned into discrete temporal modes. The choice of bin size involves a tradeoff between temporal misassignment and accidental coincidences. For bin sizes comparable to the timing uncertainty of the retrieved photon, photons may be incorrectly assigned to neighboring temporal modes. Conversely, excessively large bins increase the probability of accidental coincidences and reduce the distinguishability of adjacent time bins.

\begin{figure}
    \centering
    \includegraphics[]{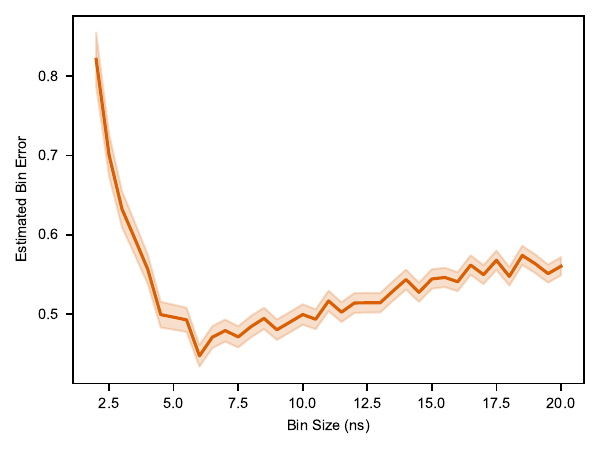}
    \caption{\textbf{Optimization of the temporal-bin size for qudit entanglement.} Estimated bin-assignment error as a function of temporal bin size. The error is determined from the reconstructed signal--idler coincidence histogram after rebinning the two-dimensional arrival-time data. A minimum error is obtained for a bin size of 6 ns, which is used for the qudit entanglement experiment in the main text. Shaded regions indicate one standard deviation assuming Poissonian statistics.
}
    \label{fig:qudit_err_analysis}
\end{figure}

To determine the optimal bin size, the coincidence histogram was rebinned over a range of temporal widths. For each bin size, the diagonal coincidence distribution was used to reconstruct the corresponding signal--idler coincidence histogram, and an effective bin error was estimated from the ratio of accidental to correlated coincidence counts. The resulting analysis is shown in Fig.~\ref{fig:qudit_err_analysis}. The estimated error exhibits a clear minimum at a bin size of 6 ns, which was subsequently used for all qudit experiments in the main text.

\end{appendices}